# THE PERFORMANCE OF LAGRANGIAN PERTURBATION SCHEMES AT HIGH RESOLUTION

by

# Thomas Buchert[1], Georgios Karakatsanis[2]
# Robert Klaffl[2], Peter Schiller[2]

[1] Theoretische Physik, Ludwig–Maximilians–Universität
Theresienstr. 37
D–80333 München
F. R. G.

[2] Max–Planck–Institut für Astrophysik
Postfach 1523
D–85740 Garching
F. R. G.

# The performance of Lagrangian perturbation schemes at high resolution

by

T. Buchert, G. Karakatsanis, R. Klaffl, P. Schiller

**Summary:** We present high–spatial resolution studies of the density field as predicted by Lagrangian perturbation approximations up to the third order. The first–order approximation is equivalent to the "Zel'dovich approximation" for the type of initial data analyzed. The study is performed for two simple models which allow studying of typical features of the clustering process in the early non–linear regime. We calculate the initial perturbation potentials as solutions of Poisson equations algebraically, and automate this calculation for a given initial random density field. The presented models may also be useful for other questions addressed to Lagrangian perturbation solutions and for the comparison of different approximation schemes. In an accompanying paper we investigate a detailed comparison with various N–body integrators using these models (Karakatsanis & Buchert 1995).
Results of the present paper include the following: 1. The collapse is accelerated significantly by the higher–order corrections confirming previous results by Moutarde *et al.* (1991); 2. the spatial structure of the density patterns predicted by the "Zel'dovich approximation" differs much from those predicted by the second– and third–order Lagrangian approximations; 3. Second–order effects amount to internal substructures such as "second generation" –pancakes, –filaments and –clusters, as are also observed in N–body simulations; 4. The third–order effect gives rise to substructuring of the secondary mass–shells. The hierarchy of shell–crossing singularities that form features small high-density clumps at the intersections of caustics which we interprete as gravitational fragmentation.

# 1. Introduction

It is commonly appreciated that Lagrangian perturbation solutions provide useful models of large–scale structure. Comparison with numerical simulations have put them into a strong position in the list of currently discussed analytical or semi–analytical approximations (see: Melott 1994 for a summary). Lagrangian perturbation schemes have been optimized by smoothing the high–frequency end of the power spectrum of density inhomogeneities such that they are capable of replacing N–body integrators above some scale close, but smaller than the non–linearity scale (i.e., where the r.m.s. density contrast is of order unity) (Coles *et al.* 1993, Melott *et al.* 1994, 1995, Bouchet *et al.* 1995, Sathyaprakhash *et al.* 1995, Weiß *et al.* 1995). While their application to pancake models, i.e., models with a large high–frequency cutoff, demonstrates an excellent performance up to the epoch when shell–crossing singularities in the cosmic flow develop (Buchert *et al.* 1994), their application to later non–linear stages fails unless the initial data are smoothed to avoid substantial post–singularity evolution. This way the large–scale structure is restored, and small–scale features arise due to the collapse of the waves which were left in the initial data.

In the present work we want to look in more detail at the collapsing structures around the epoch of shell–crossing on smaller scales by using high–resolution techniques described by Buchert & Bartelmann (1991) (however, here, we do not interpolate initial data). The present study can be viewed in line with previous high–resolution studies of pancakes (Buchert 1989a,b, Melott & Shandarin 1990, Beacom *et al.* 1991 (2D), and Buchert & Bartelmann 1991 (2D and 3D)).

The applicability of the Lagrangian approximations has been tested in previous work on the basis of cross–correlation statistics of density fields in which the internal substructures are not resolved. We here address the question which substructures are predicted by these approximations and we shall single out first–, second–, and third–order effects in the evolution of caustics in the density field. The work by Alimi *et al.* (1990), Moutarde *et al.* (1991) and the comparison of Lagrangian perturbation solutions with the spherically symmetric solution by Munshi *et al.* (1994) comprise steps in this direction.

We have taken care of the precision with which we realize the Lagrangian schemes. Thus far, these analytical models have to be realized numerically to set up the initial data in Fourier space. In particular, the third–order model provides a complication, since products of derivatives of perturbation potentials at first and second order (as solutions of Poisson equations) form the input for the third–order perturbation potentials, which describe interaction of perturbations (see the next subsection for details). It is therefore desirable to control this realization of initial data in an optimal way to minimize numerical uncertainties. We did this by calculating the perturbations fully analytically. We have also automated the process of finding the perturbation potentials from a single given initial velocity potential, or density field, respectively, by using algebraic manipulation systems. This procedure is suitable for spectra with not too much modes like in models with coherence length. Another advantage of this analytical procedure is the possibility of improving particle number, since we are not limited by storage as in the case of FFT realizations. We therefore can present realizations using $1024^3$ particles. The structures shown can only be seen at resolutions higher than $256^3$ particles.

We start with the derivation of a simple plane–wave model attempted earlier by Moutarde *et al.* (1991) (see also Alimi *et al.* 1990). Their third–order solution was derived just for this model. (However, this solution did not pass a test we did by inserting it into the Euler–Poisson system in Lagrangian form.) We, here, proceed differently. We start from the generic third–order solution given by Buchert (1994) and insert the plane–wave model as a special case. Since both the generic model and the special model have been checked to solve the original equations (by using algebraic manipulation systems), we are confident in our calculations. Besides automating algebraically the derivation of the potentials as mentioned above, we also exemplify the use of a set of local forms given by Buchert & Ehlers (1993) and Buchert (1994) in the case of the Moutarde *et al.* problem (see the APPENDIX). We stick to that model first, since it is simple and already shows the principal features of the gravitational collapse we are interested in. Also in other work on related subjects this model is useful as an example (Mo & Buchert 1990, Matarrese *et al.* 1992), and can be used as a toy–model to compare different approximation schemes. We then move to generic initial data, i.e., data with no symmetry, but restricted to a small enough box to assure the resolution of patterns we are interested in.

## 2. A generic third–order solution

Let us recall the class of third–order solutions on which we base our models. We require that, initially, the peculiar-velocity $\vec{u}(\vec{X}, t)$ to be proportional to the peculiar-acceleration $\vec{w}(\vec{X}, t)$:

$$\vec{u}(\vec{X}, t_0) = \vec{w}(\vec{X}, t_0) t_0 \quad , \tag{1}$$

where we have defined the fields as usual (compare Peebles 1980, Buchert 1992). Henceforth, we denote the peculiar–velocity potential at the initial time $t_0$ by $\mathcal{S}$, $\vec{u}(\vec{X}, t_0) =: \nabla_0 \mathcal{S}$, and the peculiar–gravitational potential at $t_0$ by $\phi$, $\vec{w}(\vec{X}, t_0) =: -\nabla_0 \phi$, where $\nabla_0$ denotes the nabla operator with respect to the Lagrangian coordinates $\vec{X}$. The restriction (1) has proved to be appropriate for the purpose of modeling large–scale structure, since the peculiar–velocity field tends to be parallel to the gravitational peculiar–field strength after some time, related to the existence of growing and decaying solutions in the linear regime (Buchert & Ehlers 1993, Buchert 1994).

With a superposition ansatz for Lagrangian perturbations of an Einstein–de Sitter background the following mapping $\vec{q} = \vec{F}(\vec{X}, a)$ as *irrotational* solution of the Euler–Poisson system in Lagrangian form up to the third order in the perturbations from homogeneity has been obtained (Buchert 1994). $\vec{F}$ defines the displacement map from Lagrangian coordinates $\vec{X}$ to Eulerian coordinates $\vec{q}$ which are comoving with the unperturbed Hubble–flow; the general set of initial conditions $(\phi(\vec{X}), \mathcal{S}(\vec{X}))$ is restricted according to $\mathcal{S} = -\phi t_0$ (see equation (1)); $a(t) = (t/t_0)^{2/3}$:

$$\vec{F} = \vec{X} + q_1(a) \nabla_0 \mathcal{S}^{(1)}(\vec{X}) + q_2(a) \nabla_0 \mathcal{S}^{(2)}(\vec{X})$$

$$+ q_3^a(a) \nabla_0 \mathcal{S}^{(3a)}(\vec{X}) + q_3^b(a) \nabla_0 \mathcal{S}^{(3b)}(\vec{X}) - q_3^c(a) \nabla_0 \times \vec{\mathcal{S}}^{(3c)}(\vec{X}) \quad , \tag{2}$$

with:

$$q_1 = \left(\frac{3}{2}\right)(a-1) \quad , \tag{2a}$$

$$q_2 = \left(\frac{3}{2}\right)^2 \left(-\frac{3}{14}a^2 + \frac{3}{5}a - \frac{1}{2} + \frac{4}{35}a^{-\frac{3}{2}}\right) \quad , \tag{2b}$$

$$q_3^a = \left(\frac{3}{2}\right)^3 \left(-\frac{1}{9}a^3 + \frac{3}{7}a^2 - \frac{3}{5}a + \frac{1}{3} - \frac{16}{315}a^{-\frac{3}{2}}\right) \quad , \tag{2c}$$

$$q_3^b = \left(\frac{3}{2}\right)^3 \left(\frac{5}{42}a^3 - \frac{33}{70}a^2 + \frac{7}{10}a - \frac{1}{2} + \frac{4}{35}a^{-\frac{1}{2}} + \frac{4}{105}a^{-\frac{3}{2}}\right) \quad , \tag{2d}$$

$$q_3^c = \left(\frac{3}{2}\right)^3 \left(\frac{1}{14}a^3 - \frac{3}{14}a^2 + \frac{1}{10}a + \frac{1}{2} - \frac{4}{7}a^{-\frac{1}{2}} + \frac{4}{35}a^{-\frac{3}{2}}\right) \quad , \tag{2e}$$

where the initial displacement vectors have to be constructed by solving seven elliptic boundary value problems (summation over repeated indices; i,j,k = 1,2,3 with cyclic ordering).

$$\Delta_0 \mathcal{S}^{(1)} = I(\mathcal{S}_{,i,k}) \, t_0 \quad , \tag{2f}$$

$$\Delta_0 \mathcal{S}^{(2)} = 2II(\mathcal{S}^{(1)}_{,i,k}) \quad , \tag{2g}$$

$$\Delta_0 \mathcal{S}^{(3a)} = 3III(\mathcal{S}^{(1)}_{,i,k}) \quad , \tag{2h}$$

$$\Delta_0 \mathcal{S}^{(3b)} = \sum_{a,b,c} \epsilon_{abc} \frac{\partial(\mathcal{S}^{(2)}_{,a}, \mathcal{S}^{(1)}_{,b}, X_c)}{\partial(X_1, X_2, X_3)} \quad . \tag{2i}$$

$$(\Delta_0 \vec{\mathcal{S}}^{(3c)})_k = \epsilon_{pq[j} \frac{\partial(\mathcal{S}^{(2)}_{,i]}, \mathcal{S}^{(1)}_{,p}, X_q)}{\partial(X_1, X_2, X_3)} \quad . \tag{2j,k,l}$$

An important remark relevant to any realization of the solution (2) concerns the possibility of setting $\mathcal{S}^{(1)} = \mathcal{S}t_0$ without loss of generality, if initial data are spatially periodic (compare Buchert 1995b, Ehlers & Buchert 1995 for details and proofs). With this setting, the first–order solution reduces to the well–known "Zel'dovich approximation" (Zel'dovich 1970, 1973; Buchert 1992), which then assumes its familiar *local* form. Also, the truncated third–order model (i.e., neglecting interaction terms) is then, although of course *non–locally*, expressible in terms of the initial data (compare eqs. (2f–h)).

The scalar potential $\mathcal{S}^{(3b)}$ and the vector potential $\vec{\mathcal{S}}^{(3c)}$ generate interaction among the first– and second–order perturbations. The general interaction term is not purely longitudinal: inspite of the irrotationality of the flow in Eulerian space, vorticity is generated in Lagrangian space starting at the third order for this set of intial data. For more general initial data, this happens already at second order. As our analysis of the solution will show, it has sense to include the interaction term $\mathcal{S}^{(3b)}$ only, neglecting the transverse part altogether. However, as will be demonstrated, keeping only the generating function $\mathcal{S}^{(3a)}$ results in a density pattern, which is not an adequate generalization of the second–order approximation. This "truncated third–order" model has been proposed in (Buchert 1994) as the "main body" of the perturbation sequence in the early nonlinear regime, since all higher–order solutions are made up of interaction terms among the perturbation potentials. A closer look at the features presented in this work shows that the third–order model *has to be* run with the interaction term $\mathcal{S}^{(3b)}$.

# 3. Special clustering models in closed form

In the APPENDIX we demonstrate how to construct special models by using "local forms" for the displacement vectors. Although analytically interesting, this procedure is cumbersome if applied to more complex initial data. In this section we describe how we can automate the process of finding closed form expressions for the perturbation potentials.

In general, we are interested in a class of initial data which can be represented by a finite Fourier sum of plane waves having random amplitudes and random phases. The random variables can, e.g., be specified in terms of a power spectrum of a Gaussian random density field. Usually, such initial conditions are generated by FFT (Fast Fourier Transform), a method which was also used to realize the generic model in (Buchert *et al.* 1994, Melott *et al.* 1995, Weiß *et al.* 1995). However, there are two limitations of this method which both restrict the power of spatial resolution, an advantage which is in principle offered by analytical solutions. One of these limitations is due to the limited CPU storage for employing the FFT routine, the other is due to a lack of precision which may arise by constructing the initially small displacements from a given density field, or by interpolating the particle displacements into a smooth density field (using, e.g., CIC binning), respectively. As an alternative, we suggest to solve the Poisson equations in eqs. (2) algebraically by comparing the coefficients of Fourier sums in the source terms and the perturbation potentials. This way the solutions can be calculated to high accuracy without hitting on CPU storage limitations. Since the model is a one–timestep mapping, the CPU time needed for the realization is still reasonably small ($512^3$ particles require CPU times of a few hours for the generic model discussed below). However, we admit that the algebraic procedure to solve for all seven perturbation potentials in (2) is still limited by the CPU storage for the algebraic program, and the compilation time of, e.g., plot routines can be large for a large number of Fourier modes. Using the manipulation system *Mathematica*, we are easily able to construct all perturbation potentials for $\approx 50$ Fourier modes on a workstation with 256M storage. The results obtained with this method have also been checked to solve the original equations using two algebraic manipulation systems (*Reduce* and *Mathematica*).

For the special models constructed algebraically in this way we have also run the previous code (using FFT), which constructs displacements from given density fields (A.G. Weiß, priv. comm.), and found as expected that the result is a slightly smoothed variant of a direct calculation pursued in the present work. At the same time, this was an independent check of the third–order program used in previous work (compare Weiß & Buchert 1993).

Besides the special model given in the APPENDIX (Model I), i.e., for the initial potential

$$\mathcal{S}_I^{(1)}(\vec{X}) := -\frac{\varepsilon}{(2\pi)^2} \left( \alpha_x \cos(2\pi X) + \alpha_y \cos(2\pi Y) + \alpha_z \cos(2\pi Z) \right) \quad , \quad \varepsilon = \frac{2}{3} \quad , \qquad (3a)$$

we have analyzed a generic model (Model II) with the following initial potential (here, the coordinates are normalized by $2\pi$):

$$\mathcal{S}_{II}^{(1)}(\vec{X}) := 0.1953 \Big[ 4.82\sin(-X-Y) + 3.95\cos(-X-Y) + 8.82\sin(-X-Z) + 2.5\cos(-X-Z)$$

$$+ 5.32\sin(-X) + 2.11\cos(-X) + 3.29\sin(-X+Z) + 1.83\cos(-X+Z) + 6.7\sin(-X+Y)$$

$$+ 4.05\cos(-X+Y) + 6.92\sin(-Y-Z) + 1.2\cos(-Y-Z) + 3.8\sin(-Y) + 4.77\cos(-Y)$$

$$+ 1.8\sin(-Y+Z) + 4.58\cos(-Y+Z) + 1.29\sin(-Z) + 6.6\cos(-Z) + 8.25\sin(Z)$$

$$+ 4.77\cos(-Z) + 3.67\sin(Y-Z) + 3.48\cos(Y-Z) + 2.4\sin(Y) + 6.02\cos(Y)$$

$$+ 7.86\sin(Y+Z) + 6.64\cos(Y+Z) + 9.33\sin(X-Y) + 0.87\cos(X-Y)$$

$$+ 0.56\sin(X-Z) + 4.48\cos(X-Z) + 3.4\sin(X) + 5.77\cos(X) + 4.54\sin(X+Z)$$

$$+ 4.46\cos(X+Z) + 9.8\sin(X+Y) + 3.13\cos(X+Y) \Big] \; . \qquad (3b)$$

The coefficients have been determined by the requirement that the power spectrum had the slope +1 down to the smallest wavelength, and the r.m.s. density contrast had the same value as Model I. Model I is the model studied by Moutarde *et al.* (1991); it has also been used by Mo & Buchert (1990) (at first order) as a statistical toy–model, and by Buchert & Ehlers (1993) (at second order) to demonstrate secondary shell–crossings; Matarrese *et al.* (1992) and Kate Croudace (priv. comm.) have compared general relativistic with Newtonian dynamics with the help of this model.

All seven perturbation potentials and the corresponding displacement vectors are listed in the APPENDIX for Model I. For Model II the potentials and the displacement vectors can be obtained on request.

# 4. High–resolution Studies

We present high–resolution studies of the density field as predicted by the Lagrangian schemes for both models. This is done by collecting $1024^3$ trajectories into a (comoving) Eulerian grid of $512^3$ cells for Model I ($512^3$ into a grid of $256^3$ for Model II) (for the method see Buchert & Bartelmann 1991).

Figure 1 displays three evolution stages of the density field predicted by Model I for the first–, second–, and third–order perturbation solutions. (Initial data were given at $z_i = 1000$; $a(z_i) = 1$.) A manifest feature is the delay of the collapse time for perturbation solutions at different orders; higher–order corrections significantly accelerate the collapse. This result was already stated by Moutarde *et al.* (1991). To compare the spatial patterns for the different orders, we can roughly compare the density fields "diagonally" in Fig.1 (this way of comparison will be discussed quantitatively in a forthcoming paper: Karakatsanis & Buchert 1995): while the first–order solution (the "Zel'dovich approximation") carries mainly kinematical information beyond the epoch of shell–crossing, the second–order solution modifies the shape of the first mass–shell after crossing and generates a second mass–shell as well as secondary sheets and filaments inside the first structures (in agreement with the previous study of the trajectory field by Buchert & Ehlers 1993); the third–order correction redistributes mass inside the two mass–shells as well as in sheets and filaments.

Figure 2 displays the third–order density field for another color coding which is useful to separate the different parts of the third–order corrections in the solution (2): we infer that the transverse part of the "interaction term" (2j,k,l) is not of crucial importance and might be neglected, it merely deconcentrates the inner mass–shell more from the center (which can be seen by comparing full third–order with or without transverse part, or "truncated third–order" with or without transverse part). However, to neglect the "interaction terms" altogether results in a pattern which is further away from the second–order approximation than the full third–order approximation. The outer caustic is even absent. This indicates that the third–order approximation without "interaction terms" is not useful, the "main body" of the perturbation sequence is not a good model as was speculated in (Buchert 1994).

We continue by looking at Figure 3 which presents the density field of Model II for the different orders at a late evolution stage, late, because we then are able to separate the different structures visually which appear much earlier in the evolution. Again the features quoted above are visible, the collapse is delayed by a huge factor in the first–order ("Zel'dovich–")approximation. Also, the similarity between second– and third–order is striking, while the first–order model lacks some internal structures, which can be attributed to secondary shell–crossing events (a second–order effect).

An interesting aspect of these high–resolution studies relates to a new interpretation of the longstanding "fragmentation problem" in classical pancake theory: we appreciate small "fragments" sitting at the intersection of caustics (see Figs.1–3). Since a finite resolution brings the density to a finite value, these "fragments" show up as almost spherical blobs with potential wells that have about 2 times more height than the potential of the mass–shells. In realistic situations, physical processes at the location of caustics and velocity dispersion in a dark collisionless component will do a similar job. We may interpret this phenomenon as "gravitational fragmentation": although the initial fluctuation is coherent like in pancake models, the collapse process forms fragments on a substantially smaller scale. This interpretation is appropriate, if gravity is the dominating interaction related to the existence of a mass dominating dark matter component in the Universe. It has far reaching consequences in a self–gravitating medium, since we expect the phenomenon of multiple shell–crossing (termed "non–dissipative gravitational turbulence by Gurevich & Zybin 1988a,b) to continue down to smaller and smaller scales yielding a hierarchy of nested caustics. This has been demonstrated in a two–dimensional simulation by Doroshkevich *et al.* (1980). Since further and further mass–shells are generated in the center of a cluster, more and more caustics are simultaneously present and consequently create a huge number of "fragments" as an internal organization of mass–shells. This way, a cluster naturally creates a gravitational potential which is distinctly rippled and thereby prepares the sites for galaxy formation: we expect the baryons to preferentially drop into these "fragments".

Although this consideration has to remain premature at this stage, we think that "gravitational fragmentation" as we describe it is a generic effect in gravitational clustering and should be taken seriously as soon as a dark matter component dominates the matter density. The fact that we need high–spatial resolution studies to uncover these fragments explains their absence in the literature. It is interesting to note here that another type of fragments appeared in a two–dimensional numerical simulation at high resolution (Melott & Shandarin 1990), which results from a redistribution of mass inside filaments (compare their plot with the filaments in the third–order approximation in Fig.1).

The detailed study of caustic metamorphoses begun by Arnol'd *et al.* (1982) for the "Zel'dovich–approximation" in two spatial dimensions will provide the necessary insight to further understand this phenomenon. We have continued this study in three spatial dimensions (Buchert *et al.* 1995a,b); for an overview see (Buchert 1995a).

**Acknowledgments:** We would like to thank Jean–Michel Alimi for discussions, and Arno Weiß for constructive contributions and discussions. GK and TB were both supported by the "Sonderforschungsbereich 375–95 für Astro–Teilchenphysik der Deutschen Forschungsgemeinschaft". All authors acknowledge financial support from a DAAD exchange program.

# Figure Captions

**Figure 1**: Three stages at expansion factors ($a = 1000, 1200, 1500$) are shown for the first–order approximation (the "Zel'dovich approximation"), (top), the second–order approximation (middle), and the third–order approximation (bottom). The initial condition is a special periodic function which maps principal elements of the large–scale structure such as sheets, filaments and clusters. At a stage shortly after the first shell–crossing (in this normalization at $a = 1000$) the three approximations mainly differ in their prediction of the collapse time. The higher–order corrections accelerate the collapse significantly and generate "second generation" pancakes, filaments and clusters. If we renormalize the amplitudes such that the collapse occurs at the same instant in all three approximations, then we can read the figures in a diagonal manner, i.e., the second stage in the upper row roughly corresponds to the first stage in the middle row, etc. .

**Figure 2**: A zoom (1/4 of the box) into the density pattern of the third–order approximation for Model I is shown with a different color coding. The contributions to the third–order effect have been splitted into a part (upper left) which belongs to the "main body"(see Section 2), a longitudinal interaction effect added to it (upper right), and a transverse interaction effect added to it (lower right). The full third–order approximation is shown in the lower left corner.

**Figure 3**: A comparison similar to **Figure 1**, however, for the generic clustering model. The density pattern predicted by the first–order ("Zel'dovich–)approximation is shown in the upper left panel, that for second–order in the upper right, for third–order in the lower left, and for third–order without "interaction terms" in the lower right panel. The same effects as quoted for Model I can be seen.

# APPENDIX

## A.I. The first–order displacement vector

In this section we exemplify how special solutions to (2) can be constructed by using a set of local forms which provide first integrals of the seven Poisson equations (2f-l). Let us consider a simple plane–wave model for the initial peculiar–velocity potential $\mathcal{S}; \mathcal{S}^{(1)} := \mathcal{S}t_0$, which was studied earlier by Moutarde *et al.* (1991):

$$\mathcal{S}^{(1)}(\vec{X}) := -\frac{\varepsilon}{(2\pi)^2} \left(\alpha_x \cos(2\pi X) + \alpha_y \cos(2\pi Y) + \alpha_z \cos(2\pi Z)\right) \quad . \tag{A.1a}$$

The amplitude $\varepsilon$ plays the role of the perturbation parameter here and is related to the total amplitude $\sigma$ of the density contrast $\delta := \frac{\rho - \rho_H}{\rho_H}$ as $\sigma = \frac{3}{2}\varepsilon$. The amplitudes $\alpha_x, \alpha_y, \alpha_z$ allow for triaxial deformations of the model; one has to choose $\alpha_x^2 + \alpha_y^2 + \alpha_z^2 = 1$ in order to keep the r.m.s. amplitude of $\delta$ the same. In this paper we shall use $\alpha_x = 1, \alpha_y = 1, \alpha_z = 1$, since different amplitudes give no further information about internal structures of the model. Although the model (A.1a) is simple, it has no symmetries which destroy the generic feature of the singularites formed like plane or spherical symmetry would do. The structure of the cluster formed will only retain reflection and rotational symmetries manifest in the potential (A.1a) for our choice of amplitudes. As, e.g., demonstrated in (Buchert & Ehlers 1993) for a similar two–dimensional model, we have with models like (A.1a) the possibility of studying principal kinematical features of a generic collapse such as the formation of cusped caustics, interconnected network structures, infall of matter onto the cluster. Additionally, internal differentiation of a multi–stream system resulting in a hierarchy of shell–crossings, which are attributed to a generic feature of a gravitational collapse, can be demonstrated nicely with this model. The model has periodic boundary conditions which makes it accessible for numerical treatment.

From (A.1a) we have for the first order displacement vector:

$$\nabla_0 \mathcal{S}^{(1)} = \frac{\varepsilon}{2\pi} \begin{pmatrix} \alpha_x \sin(2\pi X) \\ \alpha_y \sin(2\pi Y) \\ \alpha_z \sin(2\pi Z) \end{pmatrix} \quad . \tag{A.1b}$$

We now scetch a procedure how to construct the higher–order potentials from this initial condition. The procedure is based on a list of *local forms* given by Buchert & Ehlers (1993) and Buchert (1994), which are, roughly speaking, first integrals of the quadratic and cubic source terms in the Poisson equations of the solution (2). These integrals only hold for special classes of initial data, although they might also be useful as approximations for generic initial data. For the potential (A.1a) it turns out that it belongs to the class of initial data which, for all orders, admits such first integrals.

## A.II. The second–order displacement vector

According to *COROLLARY 1* proved in (Buchert & Ehlers 1993), a local form can be obtained for second order displacements. It reads

$$\nabla_0 \mathcal{S}^{(2)} = \nabla_0 \mathcal{S} \left(\Delta_0 \mathcal{S}\right) - \left(\nabla_0 \mathcal{S} \cdot \nabla_0\right) \nabla_0 \mathcal{S} \; ; \; \nabla_0 \mathcal{S} \times \Delta_0 \nabla_0 \mathcal{S} = \vec{0} \quad . \tag{A.2a,b,c,d}$$

The local form (A.2) is constructed such that its divergence agrees with the source term in (2g), its curl is, however, in general non-zero, it only vanishes if (A.2b,c,d) are statisfied. Inserting the potential (A.1a) we immediately obtain the second–order displacement vector:

$$\nabla_0 \mathcal{S}^{(2)} = \frac{\varepsilon^2}{2\pi} \begin{pmatrix} \alpha_x \sin(2\pi X)(\alpha_y \cos(2\pi Y) + \alpha_z \cos(2\pi Z)) \\ \alpha_y \sin(2\pi Y)(\alpha_x \cos(2\pi X) + \alpha_z \cos(2\pi Z)) \\ \alpha_z \sin(2\pi Z)(\alpha_x \cos(2\pi X) + \alpha_y \cos(2\pi Y)) \end{pmatrix} \quad . \tag{A.2e}$$

The vector (A.2e) is curl–free as can be easily demonstrated, so it obeys the constraints (A.2b,c,d) necessary to admit a potential. This potential can now be guessed from (A.2e) to be of the form

$$\mathcal{S}^{(2)} := -\frac{\varepsilon^2}{(2\pi)^2} \left( \alpha_x \alpha_y \cos(2\pi X) \cos(2\pi Y) + \alpha_y \alpha_z \cos(2\pi Y) \cos(2\pi Z) \right.$$

$$\left. + \alpha_x \alpha_z \cos(2\pi X) \cos(2\pi Z) \right) \; . \tag{A.2f}$$

## A.III. The third–order displacement vector of the "truncated model"

Similarily, we can ask for a local vector form whose divergence agrees with the source term in equation (2h). An expression given in Buchert (1994, COROLLARY 1) has the required property: The vector $\nabla_0 \mathcal{S}^{(3a)}$ with the components

$$(\nabla_0 \mathcal{S}^{(3a)})_k = \sum_i (\nabla_0 \mathcal{S}^{(1)})_{,i} J^S_{i,k} \tag{A.3a}$$

has the property

$$\Delta_0 \mathcal{S}^{(3a)} = 3III(\mathcal{S}^{(1)}_{,i,k}) \; ,$$

where $J^S_{i,k}$ are the subdeterminants of the tensor $(\mathcal{S}^{(1)}_{,i,k})$ (a comma always denotes partial derivative with respect to Lagrangian coordinates). The following constraints have to be satisfied in order that $\nabla_0 \mathcal{S}^{(3a)}$ be curl–free:

$$\sum_i (\nabla_0 \mathcal{S}^{(1)})_{,i} J^S_{i,[k,j]} = 0 \; , \quad k \neq j \; . \tag{A.3b,c,d}$$

Inserting the potential (A.1a) into (A.3a) gives for the displacement vector

$$\nabla_0 \mathcal{S}^{(3a)} = \frac{\varepsilon^3}{2\pi} \alpha_x \alpha_y \alpha_z \begin{pmatrix} \sin(2\pi X) \cos(2\pi Y) \cos(2\pi Z) \\ \sin(2\pi Y) \cos(2\pi X) \cos(2\pi Z) \\ \sin(2\pi Z) \cos(2\pi X) \cos(2\pi Y) \end{pmatrix} \; . \tag{A.3e}$$

Again, the vector (A.3e) is found to be curl–free which renders the contraints (A.3b,c,d) satisfied. A potential generating this displacement is again easily found from (A.3e). It reads

$$\mathcal{S}^{(3a)} := -\frac{\varepsilon^3}{(2\pi)^2} \alpha_x \alpha_y \alpha_z \left( \cos(2\pi X) \cos(2\pi Y) \cos(2\pi Z) \right) \; . \tag{A.3f}$$

## A.IV. The third–order displacement vector of the interaction term – longitudinal part

The source term in (2i) which describes the longitudinal part of the interaction of first– and second–order perturbations has a similar structure as the second–order source term (2g). We are able to construct a local form by analogy (Buchert & Ehlers 1993, COROLLARY 1):

$$\nabla_0 \mathcal{S}^{(3b)} =$$

$$\lambda_1 \left( \nabla_0 \mathcal{S}^{(2)} (\Delta_0 \mathcal{S}^{(1)}) - (\nabla_0 \mathcal{S}^{(2)} \cdot \nabla_0) \nabla_0 \mathcal{S}^{(1)} \right) + \lambda_2 \left( \nabla_0 \mathcal{S}^{(1)} (\Delta_0 \mathcal{S}^{(2)} - \nabla_0 \mathcal{S}^{(1)} \cdot \nabla_0) \nabla_0 \mathcal{S}^{(2)} \right) \; . \tag{A.4a}$$

Here, the linear combination of the two possible integrals as a general integral has to be taken, where $\lambda_1 + \lambda_2 = 1$. In order to satisfy the requirement that the vector (A.4a) be a solution of the Poisson equation (2i), we have to assure that it is curl–free which implies (BE93, COROLLARY 1):

$$\lambda_1 \left( \nabla_0 \mathcal{S}^{(2)} \times \Delta_0 \nabla_0 \mathcal{S}^{(1)} \right) + \lambda_2 \left( \nabla_0 \mathcal{S}^{(1)} \times \Delta_0 \nabla_0 \mathcal{S}^{(2)} \right) = \vec{0} \; . \tag{A.4b,c,d}$$

As can be seen from (A4), we have to determine the parameters $\lambda_1$ and $\lambda_2$ suitably in order to fulfil the constraints (A.4b,c,d). Although, we can find the two first integrals for the potential (A.1a), the resulting vectors are not curl–free. It is a matter of some algebra until one finds the correct linear combination of the two vectors, which is curl–free. This can be achieved by first guessing the form of the potential $\mathcal{S}^{(3b)}$ from the two vectors. It is clear that, in general, we will not be successful. We obtain $\lambda_1 = \frac{3}{5}$, $\lambda_2 = \frac{2}{5}$. Thus, the displacement vector reads

$$\nabla_0 \mathcal{S}^{(3b)} = \frac{\varepsilon^3}{2\pi} \left\{ \frac{3}{5} \begin{pmatrix} \alpha_x \sin(2\pi X)\{\alpha_y \cos(2\pi Y) + \alpha_z \cos(2\pi Z)\}^2 \\ \alpha_y \sin(2\pi Y)[\alpha_x \cos(2\pi X) + \alpha_z \cos(2\pi Z)]^2 \\ \alpha_z \sin(2\pi Z)[\alpha_x \cos(2\pi X) + \alpha_y \cos(2\pi Y)]^2 \end{pmatrix} + \frac{2}{5} \cdot \right.$$

$$\begin{pmatrix} \alpha_x \sin(2\pi X)[\alpha_x \cos(2\pi X)(\alpha_y \cos(2\pi Y) + \alpha_z \cos(2\pi Z)) - [\alpha_y \cos(2\pi Y) - \alpha_z \cos(2\pi Z)]^2 + (\alpha_y^2 + \alpha_z^2)] \\ \alpha_y \sin(2\pi Y)[\alpha_y \cos(2\pi Y)(\alpha_x \cos(2\pi X) + \alpha_z \cos(2\pi Z)) - [\alpha_x \cos(2\pi X) - \alpha_z \cos(2\pi Z)]^2 + (\alpha_x^2 + \alpha_z^2)] \\ \alpha_z \sin(2\pi Z)[\alpha_z \cos(2\pi Z)(\alpha_x \cos(2\pi X) + \alpha_y \cos(2\pi Y)) - [\alpha_x \cos(2\pi X) - \alpha_y \cos(2\pi Y)]^2 + (\alpha_x^2 + \alpha_y^2)] \end{pmatrix} \right\}$$

$$= \frac{\varepsilon^3}{2\pi} \frac{1}{5} \left\{ \begin{pmatrix} \alpha_x \sin(2\pi X)[2\alpha_x \cos(2\pi X)(\alpha_y \cos(2\pi Y) + \alpha_z \cos(2\pi Z)) + 10\alpha_y \alpha_z \cos(2\pi Y) \cos(2\pi Z)] \\ \alpha_y \sin(2\pi Y)[2\alpha_y \cos(2\pi Y)(\alpha_x \cos(2\pi X) + \alpha_z \cos(2\pi Z)) + 10\alpha_x \alpha_z \cos(2\pi X) \cos(2\pi Z)] \\ \alpha_z \sin(2\pi Z)[2\alpha_z \cos(2\pi Z)(\alpha_x \cos(2\pi X) + \alpha_y \cos(2\pi Y)) + 10\alpha_x \alpha_y \cos(2\pi X) \cos(2\pi Y)] \end{pmatrix} \right.$$

$$\left. + \begin{pmatrix} \alpha_x \sin(2\pi X)[\alpha_y^2 \cos^2(2\pi Y) + \alpha_z^2 \cos^2(2\pi Z) + 2(\alpha_y^2 + \alpha_z^2)] \\ \alpha_y \sin(2\pi Y)[\alpha_x^2 \cos^2(2\pi X) + \alpha_z^2 \cos^2(2\pi Z) + 2(\alpha_x^2 + \alpha_z^2)] \\ \alpha_z \sin(2\pi Z)[\alpha_x^2 \cos^2(2\pi X) + \alpha_y^2 \cos^2(2\pi Y) + 2(\alpha_x^2 + \alpha_y^2)] \end{pmatrix} \right\} , \qquad (A.4e)$$

with the potential

$$\mathcal{S}^{(3b)} := -\frac{\varepsilon^3}{(2\pi)^2} \frac{1}{5} \left\{ \alpha_x^2 \cos^2(2\pi X)[\alpha_y \cos(2\pi Y) + \alpha_z \cos(2\pi Z)] + 10\alpha_x \alpha_y \alpha_z \cos(2\pi X) \cos(2\pi Y) \cos(2\pi Z) \right.$$

$$+ \alpha_y^2 \cos^2(2\pi Y)[\alpha_x \cos(2\pi X) + \alpha_z \cos(2\pi Z)] + \alpha_z^2 \cos^2(2\pi Z)[\alpha_x \cos(2\pi X) + \alpha_y \cos(2\pi Y)]$$

$$\left. +2[\alpha_x(\alpha_y^2 + \alpha_z^2) \cos(2\pi X) + \alpha_y(\alpha_x^2 + \alpha_z^2) \cos(2\pi Y) + \alpha_z(\alpha_x^2 + \alpha_y^2) \cos(2\pi Z)] \right\} . \qquad (A.4f)$$

## A.V. The third–order displacement vector of the interaction term – transverse part

Finally, we ask for a first integral of the transverse part of the interaction vector (2j,k,l). In (Buchert 1994, COROLLARY 2) the vector form needed has been given again as a linear combination of the two possible integrals. The vector

$$\vec{\Xi} := -\nabla_0 \times \vec{\mathcal{S}}^{(3c)} =$$

$$\mu_1 \left( (\nabla_0 \mathcal{S}^{(2)} \cdot \nabla_0) \nabla_0 \mathcal{S}^{(1)} \right) - \mu_2 \left( (\nabla_0 \mathcal{S}^{(1)} \cdot \nabla_0) \nabla_0 \mathcal{S}^{(2)} \right) \qquad (A.5a,b,c)$$

has the property

$$(\nabla_0 \times \vec{\Xi})_k = \epsilon_{pq[j} \frac{\partial(\mathcal{S}^{(2)}_{,i]}, \mathcal{S}^{(1)}_{,p}, X_q)}{\partial(X_1, X_2, X_3)} , \; i,j,k = 1,2,3 \; (cyclic) \; .$$

We have to assure $\mu_1 + \mu_2 = 1$. In order to satisfy the requirement that the vector components (A.5a,b,c) be solutions of the Poisson equations (2j,k,l), we have to guarantee that the vector field $\vec{\Xi}$ is source–free which implies after using well–known vector identities

$$\Delta_0 (\nabla_0 \mathcal{S}^{(1)} \nabla_0 \mathcal{S}^{(2)}) +$$

$$\mu_1 \left( \nabla_0 \mathcal{S}^{(2)} \Delta_0 \nabla_0 \mathcal{S}^{(1)} - \nabla_0 \mathcal{S}^{(1)} \Delta_0 \nabla_0 \mathcal{S}^{(2)} \right) + \mu_2 \left( \nabla_0 \mathcal{S}^{(2)} \Delta_0 \nabla_0 \mathcal{S}^{(1)} - \nabla_0 \mathcal{S}^{(1)} \Delta_0 \nabla_0 \mathcal{S}^{(2)} \right) = 0 \; . \qquad (A.5d)$$

Again, we find the two integrals after inserting the potential (A.1a) to be not source–free. We have to determine the correct linear combination of the two integrals. As in the longitudinal case we first guess the form of the vector potential $\vec{\mathcal{S}}^{(3c)}$ from the two integrals. We are successful with the parameters $\mu_1 = \frac{3}{5}$, and $\mu_2 = \frac{2}{5}$, but we have to add another function $\mu_P \nabla_0 \mathcal{P}^{(1)}; \mu_P = \frac{1}{5}$ such that the total displacement is source–free. The potential $\mathcal{P}^{(1)}$ is given by

$$\mathcal{P}^{(1)} = \alpha_x(\alpha_y^2 + \alpha_z^2) \cos(2\pi X) + \alpha_y(\alpha_x^2 + \alpha_z^2) \cos(2\pi Y) + \alpha_z(\alpha_x^2 + \alpha_y^2) \cos(2\pi Z) \; . \qquad (A.5e)$$

(This is possible, since the local forms discussed above are only determined up to the gradient of some potential, see Buchert 1994). The vector displacement $\vec{\Xi} = -\nabla_0 \times \vec{\mathcal{S}}^{(3c)}$ reads

$$\vec{\Xi} = \frac{\varepsilon^3}{2\pi} \left\{ \frac{3}{5} \begin{pmatrix} \alpha_x^2 \sin(2\pi X) \cos(2\pi X)[\alpha_y \cos(2\pi Y) + \alpha_z \cos(2\pi Z)] \\ \alpha_y^2 \sin(2\pi Y) \cos(2\pi Y)[\alpha_x \cos(2\pi X) + \alpha_z \cos(2\pi Z)] \\ \alpha_z^2 \sin(2\pi Z) \cos(2\pi Z)[\alpha_x \cos(2\pi X) + \alpha_y \cos(2\pi Y)] \end{pmatrix} - \frac{2}{5} \cdot$$

$$\begin{pmatrix} \alpha_x^2 \sin(2\pi X)[\cos(2\pi X)(\alpha_y \cos(2\pi Y) + \alpha_z \cos(2\pi Z)) + \alpha_y^2 \cos^2(2\pi Y) + \alpha_z^2 \cos^2(2\pi Z) - (\alpha_y^2 + \alpha_z^2)] \\ \alpha_y^2 \sin(2\pi Y)[\cos(2\pi Y)(\alpha_x \cos(2\pi X) + \alpha_z \cos(2\pi Z)) + \alpha_x^2 \cos^2(2\pi X) + \alpha_z^2 \cos^2(2\pi Z) - (\alpha_x^2 + \alpha_z^2)] \\ \alpha_z^2 \sin(2\pi Z)[\cos(2\pi Z)(\alpha_x \cos(2\pi X) + \alpha_y \cos(2\pi Y)) + \alpha_x^2 \cos^2(2\pi X) + \alpha_y^2 \cos^2(2\pi Y) - (\alpha_x^2 + \alpha_y^2)] \end{pmatrix} \right\}$$

$$- \frac{\varepsilon^3}{2\pi} \frac{1}{5} \begin{pmatrix} \alpha_x(\alpha_y^2 + \alpha_z^2) \sin(2\pi X) \\ \alpha_y(\alpha_x^2 + \alpha_z^2) \sin(2\pi Y) \\ \alpha_z(\alpha_x^2 + \alpha_y^2) \sin(2\pi Z) \end{pmatrix} = \frac{\varepsilon^3}{2\pi} \frac{1}{5} \cdot$$

$$\begin{pmatrix} \alpha_x^2 \sin(2\pi X)[\cos(2\pi X)(\alpha_y \cos(2\pi Y) + \alpha_z \cos(2\pi Z)) - 2\alpha_y^2 \cos^2(2\pi Y) - 2\alpha_z^2 \cos^2(2\pi Z) + (\alpha_y^2 + \alpha_z^2)] \\ \alpha_y^2 \sin(2\pi Y)[\cos(2\pi Y)(\alpha_x \cos(2\pi X) + \alpha_z \cos(2\pi Z)) - 2\alpha_x^2 \cos^2(2\pi X) - 2\alpha_z^2 \cos^2(2\pi Z) + (\alpha_x^2 + \alpha_z^2)] \\ \alpha_z^2 \sin(2\pi Z)[\cos(2\pi Z)(\alpha_x \cos(2\pi X) + \alpha_y \cos(2\pi Y)) - 2\alpha_x^2 \cos^2(2\pi X) - 2\alpha_y^2 \cos^2(2\pi Y) + (\alpha_x^2 + \alpha_y^2)] \end{pmatrix} ,$$
$$(A.5f)$$

with the vector–potential

$$\vec{\mathcal{S}}^{(3c)} := -\frac{\varepsilon^3}{(2\pi)^2} \frac{1}{5} \begin{pmatrix} \alpha_y \alpha_z \sin(2\pi Y) \sin(2\pi Z)(\alpha_y \cos(2\pi Y) - \alpha_z \cos(2\pi Z)) \\ \alpha_x \alpha_z \sin(2\pi X) \sin(2\pi Z)(\alpha_z \cos(2\pi Z) - \alpha_x \cos(2\pi X)) \\ \alpha_x \alpha_y \sin(2\pi X) \sin(2\pi Y)(\alpha_x \cos(2\pi X) - \alpha_y \cos(2\pi Y)) \end{pmatrix} \; . \qquad (A.5g)$$

## A.VI. Remarks

For the following discussion we need the explicit expressions of the source terms in the solution (2) for the special model (A.1a). We derive

$$I(\mathcal{S}_{,i,k}) = \varepsilon \{ \alpha_x \cos(2\pi X) + \alpha_y \cos(2\pi Y) + \alpha_z \cos(2\pi Z) \} \; , \qquad (A.6a)$$

$$II(\mathcal{S}_{,i,k}) = \varepsilon^2 \{ \alpha_x \alpha_y \cos(2\pi X) \cos(2\pi Y) + \alpha_y \alpha_z \cos(2\pi Y) \cos(2\pi Z) + \alpha_x \alpha_z \cos(2\pi X) \cos(2\pi Z) \} \; , \qquad (A.6b)$$

$$III(\mathcal{S}_{,i,k}) = \varepsilon^3 \{ \alpha_x \alpha_y \alpha_z \cos(2\pi X) \cos(2\pi Y) \cos(2\pi Z) \} \; , \qquad (A.6c)$$

$$\sum_{a,b,c} \epsilon_{abc} \frac{\partial(\mathcal{S}_{,a}^{(2)}, \mathcal{S}_{,b}, X_c)}{\partial(X_1, X_2, X_3)} = \varepsilon^3 \{ \alpha_x^2 \cos^2(2\pi X)[\alpha_y \cos(2\pi Y) + \alpha_z \cos(2\pi Z)]$$

$$+ \alpha_y^2 \cos^2(2\pi Y)[\alpha_x \cos(2\pi X) + \alpha_z \cos(2\pi Z)] + \alpha_z^2 \cos^2(2\pi Z)[\alpha_x \cos(2\pi X) + \alpha_y \cos(2\pi Y)]$$

$$+ 6 \alpha_x \alpha_y \alpha_z \cos(2\pi X) \cos(2\pi Y) \cos(2\pi Z) \} \; , \qquad (A.6d)$$

$$\epsilon_{pq[j} \frac{\partial(\mathcal{S}^{(2)}_{,i]}, \mathcal{S}_{,p}, X_q)}{\partial(X_1, X_2, X_3)} = \varepsilon^3 \begin{pmatrix} \alpha_y \alpha_z \sin(2\pi Y) \sin(2\pi Z)(\alpha_y \cos(2\pi Y) - \alpha_z \cos(2\pi Z)) \\ \alpha_x \alpha_z \sin(2\pi X) \sin(2\pi Z)(\alpha_z \cos(2\pi Z) - \alpha_x \cos(2\pi X)) \\ \alpha_x \alpha_y \sin(2\pi X) \sin(2\pi Y)(\alpha_x \cos(2\pi X) - \alpha_y \cos(2\pi Y)) \end{pmatrix} \quad . \qquad (A.6e, f, g)$$

From the generating functions constructed above we infer the following property: except for the longitudinal part of the interaction term, the perturbation potentials obey equations which are typical for bound systems:

$$\Delta_0 \mathcal{S}^{(1)} = -(2\pi)^2 \, \mathcal{S}^{(1)} \; , \qquad (A.7a)$$

$$\Delta_0 \mathcal{S}^{(2)} = -(2\pi)^2 \, 2 \, \mathcal{S}^{(2)} \; , \qquad (A.7b)$$

$$\Delta_0 \mathcal{S}^{(3a)} = -(2\pi)^2 \, 3 \, \mathcal{S}^{(3a)} \; , \qquad (A.7c)$$

$$\Delta_0 \mathcal{S}^{(3b)} = -(2\pi)^2 \, \{5 \, \mathcal{S}^{(3b)} - 4 \, \mathcal{S}^{(3a)} + 2 \, \mathcal{P}^{(1)}\} \; , \qquad (A.7d)$$

$$\Delta_0 \vec{\mathcal{S}}^{(3c)} = -(2\pi)^2 \, 5 \, \vec{\mathcal{S}}^{(3c)} \; . \qquad (A.7e)$$

Recall that the condition (A.7a) implies $\nabla_0 \delta(t_0) \propto \vec{u}$ for an initially irrotational peculiar–velocity field $\vec{u}(t_0)$; $\nabla_0 \times \vec{u}(t_0) = \vec{0}$, i.e., the motion is initiated to follow the gradient of the density–contrast field. At the third order the evolution model shows that this property of the flow is lost.

(The algebraic program we used to compute the perturbation potentials for Model II also reproduces the perturbation potentials derived here.)